\documentclass[twocolumn,prb,superscriptaddress,nofootinbib]{revtex4}
\usepackage{CJK}
\usepackage[english]{babel}
\usepackage[ansinew]{inputenc}
\usepackage{boldline,multirow,xcolor,colortbl}
\usepackage{times}
\usepackage{graphicx}
\usepackage{graphics}
\usepackage{amsmath}
\usepackage{amsfonts}
\usepackage{amssymb}
\usepackage{amsbsy}
\usepackage{dsfont}
\usepackage{epstopdf}
\usepackage{makeidx}
\usepackage{subfigure}
\usepackage{color} 
\usepackage{pgf}
\usepackage{bm}
\usepackage{tikz} 
\usepackage[normalem]{ulem}
\usepackage{hyperref}
\newcommand{\bra}{\langle}
\newcommand{\ket}{\rangle}
\newcommand{\be}{\begin{equation}}
\newcommand{\ee}{\end{equation}}
\newcommand{\bea}{\begin{eqnarray}}
\newcommand{\eea}{\end{eqnarray}}

\newcommand{\ben}{\begin{eqnarray}}
\newcommand{\een}{\end{eqnarray}}

\setcitestyle{open={[},close={]}}

\makeindex
\begin{document}
\begin{CJK*}{GB}{}
\title{Emerging nonlinear Hall effect in Kane-Mele two-dimensional topological insulators}

\author{Rajesh K. Malla}
\author{Avadh Saxena}
\affiliation{Theoretical Division and Center for Nonlinear Studies, Los Alamos National Laboratory, Los Alamos, New Mexico 87545, USA}

\author{Wilton J. M. Kort-Kamp}
\affiliation{Theoretical Division, Los Alamos National Laboratory, Los Alamos, New Mexico 87545, USA}

\begin{abstract}

The recent observations of nonlinear Hall effect in time-reversal symmetry protected systems and on the surface of  three-dimensional topological insulators due to an in-plane magnetic field have attracted immense experimental and theoretical investigations in two-dimensional transition metal dichalcogenides and  Weyl  semimetals. The origin of this type of second order effect has been attributed to the emergence of a Berry curvature dipole, which requires a low-symmetry environment. Here, we propose a mechanism for generating such a second order nonlinear Hall effect in Kane-Mele two-dimensional topological insulators due to spatial and time reversal symmetry breaking in the presence of Zeeman and Rashba couplings. By actively tuning the energy gaps with external electromagnetic fields we also demonstrate that the nonlinear Hall effect shows remarkable signatures of topological phase transitions existing in the considered two-dimensional systems.
\end{abstract}

\maketitle
\end{CJK*}
Two-dimensional materials with strong intrinsic spin-orbit coupling (SOC, $\lambda_{SO}$) have attracted intense investigations due to the possibility to control and manipulate both electric and spin currents for application in various optoelectronic and spintronic devices.  One of the intriguing phenomena emerging from the SOC is  the quantum spin Hall  effect, which describes a nonzero transverse spin current
in systems with time-reversal (TRS) and spatial-inversion (SIS) symmetries \cite{KaneMele1,KaneMele2}. It has been experimentally verified  in  various semiconducting materials, including HgTe \cite{HgTe1,HgTe2} and  InAs/GaSb quantum wells \cite{InAsGaSb1,InAsGaSb2}, as well as in  WTe$_2$ transition metal dichalcogenides (TMD) \cite{WTe1,WTe2}. The inclusion of Rashba spin-orbit coupling ($\lambda_{R}$) breaks SIS in these systems and destroys  spin angular momentum conservation while the gapless edge states and the spin topological invariant (i.e., spin Chern number) remain unchanged provided $\lambda_R < \lambda_{SO}$ \cite{SpinChern}. Moreover, since the robustness of the spin Chern number is fully determined by the bulk band gap, it has been shown that the quantum spin Hall phase persists even when TRS is broken \cite{robustspinchern}. This further allows one to extend spin manipulation to TRS-broken quantum spin Hall insulator  (QSHI) states, e.g., via exchange interaction within the Kane-Mele Hamiltonian \cite{TRSbrokenQSH}. Nevertheless, studies on the optical properties and electronic transport in these cases have been largely focused on  the linear response regime \cite{Tunablebandgap1,Tunablebandgap2, WiltonPRB, WiltonPRL, WiltonNonlocal, Circular3, Wiltonnature, Tabert2014, Farias2018,Malla2018,Wu2018, Wu2020}.

Second order nonlinear electromagnetic phenomena, including second harmonic generation (SHG), sum- and difference frequency
generation,  optical rectification, and
the Pockels effect  are very sensitive to the symmetry of a crystal. In a purely semiclassical scenario, periodic systems protected by either TRS or SIS do not show any second order electric current \cite{Ezawanoncentrosymmetric}. However, it has been predicted that a second order Hall current can emerge in noncentrosymmetric TRS invariant systems due to a nonzero Berry curvature dipole moment \cite{TRS1, Sodemann}. This effect has been theoretically investigated in various TRS invariant systems \cite{TRS2,TRS3,TRS4,TRS5,TRS6,TRS7,TRS8,lahiri2021berry} and was experimentally verified in bilayer WTe$_2$ \cite{TRSexp}. In TRS protected systems the nonlinear Hall effect emerges due to the presence of a non-zero Berry curvature dipole, which requires a low-symmetry environment.  A non-zero dipole has been achieved  as a result of either the tilt of a massive Dirac cone \cite{Sodemann, Nandy2019, Xie2019}, engineering of a finite strain \cite{TRS4}, or a hexagonal warping effect \cite{TRS8}. Recently, disorder induced nonlinear Hall effect was demonstrated in a TRS protected system \cite{Quantnonlinear, Niu2019}. On the other hand, a nonlinear planar Hall effect was experimentally realized on the surface of a three-dimensional nonmagnetic topological insulator, Bi$_2$Se$_3$, with broken TRS due to an in-plane magnetic field  \cite{NPHE2019}. The spin-momentum locking near the surface together with the hexagonal warping is responsible for such a nonlinear response \cite{zhang2018theory}. In 3D materials, such as Weyl and Dirac semimetals, the origin of nonlinear planar Hall effect has been attributed to chiral anomaly \cite{Zhang2021,nandy2021chiral}, which has been extensively studied in the case of linear planar Hall effect \cite{NPHE2020,Canomaly0,Canomaly00,Canomaly000,Canomaly1,Canomaly2,Canomaly3,Canomaly4,Canomaly5,Canomaly6}. Recently, a signature of second order charge-to-spin conversion due to an unclear mechanism was reported in transition metal dichalcogenide (TMD)-graphene heterostructures supporting the Rashba-Edelstein and spin Hall effects \cite{Nanolett}.


Here, we propose a mechanism for generating second order charge, spin, and valley Hall currents in Kane-Mele type 2D topological materials emerging from the ``cross-talk" between Rashba spin-orbit coupling and in-plane magnetic field-induced Zeeman effect (Fig. 1).  We show that the Rashba-Zeeman interplay creates the necessary asymmetries in space and time to generate such second order nonlinear responses, and examine how it affects the energy band structure, the Berry curvature, and the ensuing electronic conductivities. The Rashba-Zeeman interplay induced transport has been previously studied in different quantum systems, including a 1D chain \cite{1D}, a spinful realization of Aubry-Andre model \cite{AA}, and in a 2D electron gas \cite{2D}. We argue that two-dimensional systems such as, for instance, silicene\cite {Silicene}, germanene \cite{Germanene}, stanene \cite{Stanene}, and plumbene \cite{Plumbene}, collectively called graphene family materials, as well as antiferromagnetic manganese chalcogenophosphates (MnPX$_3$, X = S, Se) \cite{Li} and perovskites \cite{Liang}, are potential platforms to verify this mechanism.  Recently a new class of families known as Jacutingaite has been shown to host the Kane-Mele type QSH phase \cite{J1,J2,J3,J4}. By tailoring the energy gaps in the Kane-Mele Hamiltonian of these systems via external interactions, e.g., using static electric fields or non-resonant circularly polarized light, one can steer the monolayers through various topological phase transitions. We demonstrate that only the Berry curvature induced second order Hall current distinguishes between various topological phases with different Chern numbers, and find a strong enhancement of Hall current near the phase transitions. 
\begin{figure}
\includegraphics[width=1.0\linewidth]{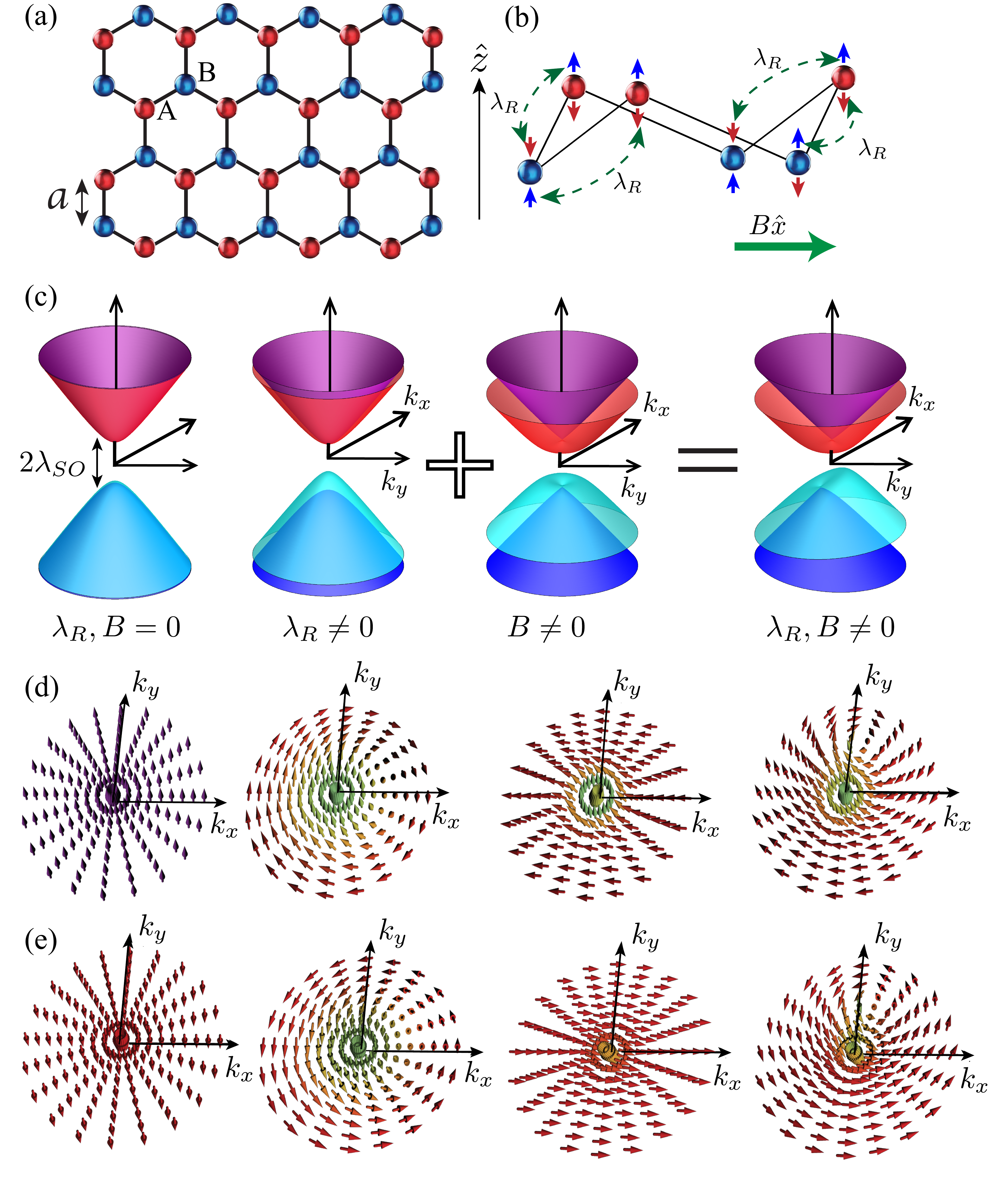}
\caption{(a) Schematic diagram of the lattice structure of hexagonal honeycomb monolayer. (b)  Representation of Rashba and Zeeman coupling. For illustration purposes we consider that the sublattices A and B have a buckled structure, characteristic of two-dimensional topological insulators such as silicene, germanane, stanene, and plumbene. (c) (Left) Energy band structure near the $K$ valley in the absence of Rashba and Zeeman coupling. (Center) Individual effects of each of these interactions in the energy-momentum dispersion. (Right) Illustration of the effect of the Rashba-Zeeman interplay in the energy band structure. (d, e) Spin-distributions of the two conduction bands in momentum space 
are shown for the cases considered in (c).}
\label{Fig1}
\end{figure}
\section{Second order nonlinear dynamics in Dirac-like 2D systems}
We consider nonlinear effects in light-matter interactions between an ac electric field and an arbitrary monolayer with unperturbed Hamiltonian ${\hat H}_0$.  The equation of motion for the density matrix operator $\hat{\rho}(t)$ reads 
\begin{equation}
i\hbar\ \frac{\partial \hat{\rho}(t)}{\partial t}= \left[{\hat H}_{0}+ \hat{H}_{i}(t),\hat{\rho}(t)\right]-\Gamma({\hat \rho(t)}-{\hat \rho}^{(0)}),
\label{EqMotRho}
\end{equation} 
where ${\hat H}_i(t)=-e\phi({\bm r},t)$ is the interaction Hamiltonian, and $\phi({\bm r},t)=\int_{-\infty}^{\infty}d\omega\sum_{\bm q} \phi_{\bm{q}\omega} e^{i ({\bm q}\cdot{\bm r}-\omega t)}$ is the scalar electromagnetic potential, which relates to the electric field as $ \boldsymbol{{\cal E}} = -\nabla \phi$. Also, ${\hat \rho}^{(0)}=\sum_{lk} f_{lk} |lk\rangle\langle lk |$ is the equilibrium density matrix. Here, $f_{lk}$ describes the Fermi-Dirac distribution (FD) of electrons with momentum $\hbar {\bm k}$ in the $l$-th band of the unperturbed Hamiltonian, and $\Gamma$ is the phenomenological decay rate. We solve Eq. \eqref{EqMotRho} within the framework of perturbation theory by expanding the time-dependent density matrix in powers of the electric field amplitude $\hat{\rho}(t)=\sum_n {\hat \rho}^{(n)}(t) \propto {\cal E}(t)^n$, and compute the $n$-th order two-dimensional current density  at ${\bm r}_0$ as ${\bm j}^{(n)}({\bm r}_0,t)=-(e/2)\sum_{ll'{\bm k}{\bm k}'}\bra l'{\bm k}'|\{{\hat {\bm v} \delta({\bm r}_0-{\bm r})\}_{+}|l{\bm k}\ket \bra l{\bm k}|\hat{\rho}^{(n)}(t)|l'{\bm k}'\ket}$, where $\{{\hat a},{\hat b}\}_{+}={\hat a}{\hat b}+{\hat b}{\hat a}$.  The velocity operator is defined as
${\hat {\bm v}}=-i[{\hat {\bm r}},{\hat H}_0]$. We consider the local limit ${\bm q}\rightarrow 0$ by neglecting spatial dispersion since we assume a normally incident electromagnetic wave, and derive expressions for the optical conductivity tensor up to second order in the perturbation expansion by enforcing that the current be proportional to the  second power of the incident field, ${\bm j}_{\alpha}=\sigma_{\alpha\beta\gamma}{\cal E}_{\beta}{\cal E}_{\gamma}$, where $\sigma_{\alpha\beta\gamma}$ is the the optical conductivity tensor describing a second order dynamics.

The generation of second order optical current involves two virtual electronic transitions between the energy bands and each transition can take place either within the same band $l=l'$ (intraband) or across two different bands $l\neq l'$ (interband). The second order optical conductivity  $\sigma_{\alpha\beta\gamma}$ can be separated into two terms ${\tilde \sigma}_{\alpha\beta\gamma}^{(1)}$ and  ${\tilde \sigma}_{\alpha\beta\gamma}^{(2)}$ that are either proportional to the FD functions or their derivatives, respectively (see appendix). The expressions for ${\tilde \sigma}_{\alpha\beta\gamma}^{(1)}$ and  ${\tilde \sigma}_{\alpha\beta\gamma}^{(2)}$ are
\begin{multline}
{\tilde \sigma}_{\alpha\beta\gamma}^{(1)}(\omega_1,\omega_2)=
\frac{e^3\hbar^2}{S}\sum\limits_{l\neq l'\neq l''{\bm k}}\frac{\bra \hat{v}_{\alpha}\ket_{l'l}}{\left(E_{l'l}+\hbar(\omega_1+\omega_2+i\Gamma)\right)}\\
\times \frac{1}{E_{ll''}E_{l''l'}}\Big[\frac{\bra \hat{v}_{\beta}\ket_{ll''}\bra \hat{v}_{\gamma}\ket_{l''l'}f_{l'l''}}{E_{l'l''}+\hbar(\omega_1+i\Gamma)}-\frac{\bra \hat{v}_{\gamma}\ket_{ll''}\bra \hat{v}_{\beta}\ket_{l''l'}f_{l''l}}{E_{l''l}+\hbar(\omega_1+i\Gamma)}\Big]\\
+
\frac{e^3\hbar}{S}\sum\limits_{l\neq l'{\bm k}}\frac{\bra \hat{v}_{\alpha}\ket_{l'l}f_{l'l}}{\left(E_{l'l}+\hbar(\omega_1+\omega_2+i\Gamma)\right)}\\
\times
\frac{\partial}{\partial k_{\gamma}}\left(\frac{\bra \hat{v}_{\beta}\ket_{ll'}}{E_{ll'
}\left(E_{l'l}+\hbar(\omega_1+i\Gamma)\right)} \right),
\label{sigma1}
\end{multline}
\begin{multline}
{\tilde \sigma}_{\alpha\beta\gamma}^{(2)}(\omega_1,\omega_2)={\tilde \sigma}_{\alpha\beta\gamma}^{(2,GV)}+{\tilde \sigma}_{\alpha\beta\gamma}^{(2,BD)}+{\tilde \sigma}_{\alpha\beta\gamma}^{(2,BDC)}=
\\
    \frac{e^3}{S}\sum\limits_{l{\bm k}}\frac{\bra{\hat v}_{\alpha}\ket_{ll}\frac{\partial^2f_{l}}{\partial k_{\beta} \partial k_{\gamma}}}{\hbar(\omega_1+\omega_2+i\Gamma)(\hbar(\omega_1+i\Gamma))}\\
    +\frac{e^3\hbar}{S}\sum\limits_{l\neq l'{\bm k}}\frac{\bra \hat{v}_{\alpha}\ket_{l'l}}{\left(E_{l'l}+\hbar(\omega_1+\omega_2+i\Gamma)\right)}\\
\times\Bigg[\frac{\bra \hat{v}_{\gamma}\ket_{ll'}\frac{\partial f_{l'l}}{\partial k_{\beta}}}{E_{ll'}\hbar(\omega_1+i\Gamma)} +
\frac{\bra \hat{v}_{\beta}\ket_{ll'}\frac{\partial f_{l'l}}{\partial k_{\gamma}}}{E_{ll'
}\left(E_{l'l}+\hbar(\omega_1+i\Gamma)\right)}
\Bigg],
\label{sigma2}
\end{multline}
where $S$ is the surface area and $\sigma_{\alpha\beta\gamma}={\tilde \sigma}_{\alpha\beta\gamma}^{(1)}+{\tilde \sigma}_{\alpha\beta\gamma}^{(2)}$ is the total conductivity, $\omega_1$ and $\omega_2$ are frequencies of the incident electric fields, and $E_{ll'}=E_{l}-E_{l'}$ with $E_l$ being the energy of the $l$th band at momentum ${\bm k}$.

Equation \eqref{sigma1} becomes important for larger frequencies and shows spectral resonances whenever a bandgap coincides with energy of either one photon $\hbar\omega_1$ or two photons $\hbar (\omega_1+\omega_2)$. We restrict our discussion to near static regime where ${\tilde \sigma}_{\alpha\beta\gamma}^{(1)}(\omega_1,\omega_2)$ becomes negligible. The first term ${\tilde \sigma}_{\alpha\beta\gamma}^{(2,GV)}$ in Eq. (\ref{sigma2})  originates from the charge carrier's group velocity (GV) $\partial E_{l}({\bm k})/\partial k_{\alpha}$, and a nonzero contribution from it requires breaking the symmetry of the energy bands $E_{l}({\bm k})\neq E_{l}(-{\bm k})$.
The second term  ${\tilde \sigma}_{\alpha\beta\gamma}^{(2,BD)}$ emerges from the nonzero dipole moment of the Berry curvature in the semiclassical (low frequency) limit $\mu\gg\hbar\omega$. The Berry dipole (BD) requires asymmetry of the Berry curvature and is given by\cite{Sodemann}
\be
D_{\alpha z}=\sum\limits_{l{\bm k}} \frac{\partial \Omega_{z,l}({\bm k})}{\partial k_{\alpha}} f_{lk},
\label{berry-dipole}
\ee
where $\alpha$ is the dipole direction and  $\Omega_{z,l}({\bm k})=2~\text{Im}[\sum_{l'\neq l}(\bra {\hat v}_{y}\ket_{ll'}\bra {\hat v}_{x}\ket_{l'l})/E_{l'l}^2]$ is the Berry curvature of $l$th band of a monolayer lying in the $xy$-plane. Note that $D_{\alpha z}$ is nonzero provided $\partial \Omega_{z,l}({\bm k})/\partial k_{\alpha} \neq -\partial \Omega_{z,l}(-{\bm k})/\partial k_{\alpha}$, which occurs in 2D materials with a single mirror symmetry line, resulting in a dipole moment of the Berry curvature in the monolayer's plane and orthogonal to the symmetry line  \cite{Sodemann}. The third term ${\tilde \sigma}_{\alpha\beta\gamma}^{(2,BDC)}$ in Eq. \eqref{sigma2} is very similar to the BD term, however, the denominator includes a third power in energy, and does not appear within the standard semiclassical approach.  This effect has been previously investigated in the dynamics of Bloch electrons under uniform electromagnetic fields and magnetically parity-violating systems \cite{Niuprl,PRXchiral} and is typically referred to as intrinsic Fermi surface effect, since it includes a derivative of the FD function. In the small frequency regime  this term is significantly smaller ($\sim \hbar \omega/\mu$) compared with the BD contribution, and therefore we call it ``Berry dipole correction" (BDC). Finally, we comment that in centrosymmetric systems the functions within the summations in Eqs. \eqref{sigma1} and \eqref{sigma2} have odd parity in momentum, leading to a vanishing second order current.

We comment that our approach can also be extended to both  spin and valley currents. 
Since, in general the spin will no longer be a good quantum number, we can use a spin-polarized velocity operator to filter the spin current along a particular spin polarization.  In order to get the spin current density, denoted by ${\bm j}_{\nu}^{(n)}({\bm r_0,t})$,  the standard velocity operator can be replaced by a spin-polarized velocity operator $ {\hat v}_{\alpha}^{(\nu)}=\{{\hat v}_{\alpha},\hat{s}_{\nu} \}/4$   in the expression for electric current density ${\bm j}^{(n)}({\bm r_0,t})$, where $\nu$ is the direction of the spin polarization and $\hat{s}$ is the spin Pauli matrix\cite{Ezawanoncentrosymmetric}. 
In the following, we will keep our discussion to spin polarization along $z$-direction, and denote the spin conductivity as $\sigma_{s,\alpha\beta\gamma}^{(2)}$. The valley conductivity is obtained by summing the contributions arising from decoupled valleys weighted by the valley index $\eta$, in a similar fashion as the spin conductivity is the sum of the conductivities from all Dirac cones weighted by the spin index $s$, $\sigma_{\eta,\alpha\beta\gamma}=\sum_{\eta} \eta \sigma_{\alpha\beta\gamma}^{(\eta)}$, where  $\sigma_{\alpha\beta\gamma}^{(\eta)}$ is the conductivity contribution from the valley $\eta$. Finally, we will also denote the spin-polarized and valley-polarized Berry curvature as $\Omega_{z,l}^{(s)}, \Omega_{z,l}^{(\eta)}$, and the spin-polarized and valley-polarized Berry curvature dipole as $D_{\alpha z}^{(s)}, D_{\alpha z}^{(\eta)}$, respectively.

\section{Low-energy Kane-Mele Hamiltonian in 2D topological insulators}
We apply the above formalism to  a Kane-Mele two-dimensional topological insulator interacting with a monochromatic ac electric field $\bm{{\cal E}}(t)$ impinging normally on the monolayer plane (Fig. 1). The low-energy unperturbed Hamiltonian of the system reads
\begin{eqnarray}
\label{DiracHamiltonian}
{\hat H}_0^{\eta s}&=&\hbar v_F(\eta k_x \tau_x +k_y \tau_y) +\Delta_{\eta}^{s}\tau_z +\frac{\lambda_{R}}{2}(\eta \tau_x s_y-\tau_y s_x)\cr
&+&\Delta_B (\cos \theta s_x+\sin \theta s_{y}),
\end{eqnarray}
where the particle momentum is measured from the Dirac points $K$  and $K'$ and $v_F$ is the Fermi velocity. Pauli matrices  $\tau_i$ correspond to sublattice pseudospins ($A$, $B$), while $s_i$ describe the electron spins.  The spin and valley dependent Dirac gap  is $\Delta_{\eta}^{s}=\eta s \lambda_{SO}$, where the  strength of the SOC $\lambda_{SO}$ depends on the buckling of the honeycomb lattice as well as the atomic size. For instance,  it is found to be $3.9$ meV for silicene,  $20$ meV for germanene, $300$ meV for stanene, and $0.4$ eV for plumbene\cite{Ezawajapanese,Wiltonnature,Plumbene}. The third term in Eq. (\ref{DiracHamiltonian}) represents the  Rashba coupling $\lambda_{R}$ (Fig. 1b), which originates from the nearest neighbor hopping in the tight-binding model and requires external parameters such as static electric field\cite{RashbaE1,RashbaE2}, metal-atom adsorption\cite{Rashbaad1,Rashbaad2}, or presence of a substrate\cite{Rashbasub}, which breaks SIS in the normal direction. Finally, we consider a Zeeman interaction via the last term in Eq. \eqref{DiracHamiltonian} by applying an in-plane magnetic field ${\bf B}$ $(\Delta_B = g_L\mu_B |B|/2)$ arbitrarily oriented along the plane of the sample with an angle $\theta$ measured from the $x$-axis (Fig. 1b). Here $g_L$ is the Lande g factor and $\mu_B$ is the Bohr magneton. Equation (\ref{DiracHamiltonian}) without the Rashba-Zeeman terms, accounts for the centrosymmetric part of the full tight-binding  KM Hamiltonian. It is protected by both  TRS, ${\cal T} {\hat H_{\eta}}({\bm k}){\cal T}^{-1}={\hat H}_{\eta}(-{\bm k})$, and SIS, ${\cal P} {\hat H_{\eta}}({\bm k}){\cal P}^{-1}={\hat H}_{-\eta}(-{\bm k})$, where the symmetry operators are given as ${\cal T}=i\gamma_0\tau_1 s_2{\cal K}$ and ${\cal P}= \gamma_1\tau_1 s_0$, and
$\gamma$, $\tau$, and $s$ are Pauli matrices associated with the valley, sublattice, and spin degrees of freedom, and ${\cal K}$ is the complex conjugation operator. In this case, the energy band-structure is degenerate in both valley and spin (Fig. \ref{Fig1}c) and the expectation value of the spin operator is orthogonal to the monolayer plane and locked to the direction of motion of the charge carriers (Fig. \ref{Fig1}d). 

The Rashba and Zeeman terms break respectively SIS and TRS and, when simultaneously applied to the monolayer, lift the the degeneracy of the energy bands $E(-{\bm k}) \neq E({\bm k})$. In the presence of a nonzero $\lambda_{R}$, the energies describing the conduction ($E_+$) and valence ($E_-$) bands  read $E_{+(-)}=\pm\lambda_{R}/2+(-)[\hbar^2v_F^2k^2)+\left(\lambda_{SO}\mp(\lambda_{R}/2) \right)^2]^{1/2}$. For momentum $k=0$, the energies of the valence bands are shifted by $-\lambda_{SO}\pm\lambda_{R}$ while the conduction bands remain unaffected, which leads to new energy gaps equal to $2|\lambda_{SO}\pm\lambda_{R}|$ see Fig. \ref{Fig1}c.   The Zeeman field affects the energy band structure as $E=\pm[\left( \hbar v_F k\pm \Delta_B\right)^2+\lambda_{SO}^2]^{1/2}$, but does not change the magnitude of the energy gaps. On the other hand, the minimum energy is shifted radially in momentum space and lies along the contour $\hbar v_F k=|\Delta_B|$, as shown in Fig. \ref{Fig1}c.   The corresponding effects for these interaction terms in the spin distribution in momentum space are shown in Fig. \ref{Fig1}d,e.  While the Rashba term results in a spin-momentum locking mechanism that leads the in-plane components of the spin to be orthogonal to the momentum, the Zeeman term tends to orient the spin parallel to the direction of the applied in-plane magnetic field. 

The independent violation of SIS or TRS via Rashba or Zeeman interaction does not break the even parity of the energy band, i. e., $E(-{\bm k})=E({\bm k})$, and therefore Rashba and Zeeman couplings cannot individually lead to any energy band dependent second order electric current. Rather, both contributions are simultaneously necessary to enable a spin-to-charge conversion mechanism that in the nonlinear regime results in an emerging nonlinear Hall current\cite{Natrev1,Ortix}. Assuming that the direction of the magnetic field is along the $x$-axis ($\theta=0$), one can show that the eigenenergies of the unperturbed Hamiltonian satisfy a quartic order equation, which includes a term proportional to $\lambda_{R}\Delta_{B} k_y$. Because this term is odd in $k_y$, it results in a band structure that does not preserve momentum inversion symmetry. Note that nonzero Zeeman and Rashba couplings are needed for this term to be relevant, further confirming that both SIS and TRS need to be broken to enable new effects in the KM model Hamiltonian. We mention that the energy dispersion along $k_x$, however, remains symmetric, $E_{l}(k_x)=E_{l}(-k_x)$. 
Note also that when both interactions are turned on, the spin distribution due to each valley is distorted in the direction orthogonal ($k_y$) to the magnetic field (Fig. 1d). This asymmetry involving spins with opposite momentum is crucial to create unbalanced spin currents, which then results in charge transport when the sample is driven by an electromagnetic wave\cite{NPHE2019}.

\section{Rashba-Zeeman enabled electronic, spin, and valley transport}

The interplay between the  Rashba coupling  $\lambda_{R}$ and Zeeman energy $\Delta_{B}$
has a strong effect on all the three terms of the conductivity tensor in Eq. (\ref{sigma2}), and we analyze each of them in detail.  We assume that the Zeeman coupling is due to a magnetic field applied along the $x$-direction, and we consider a monochromatic incident field of frequency $\omega$ and focus on second harmonic generation processes, {\it i.e.}, $\omega_1 = \omega_2 = \omega$. 
\subsection{Group velocity contribution}

\begin{figure}
\includegraphics[width=1.0\linewidth]{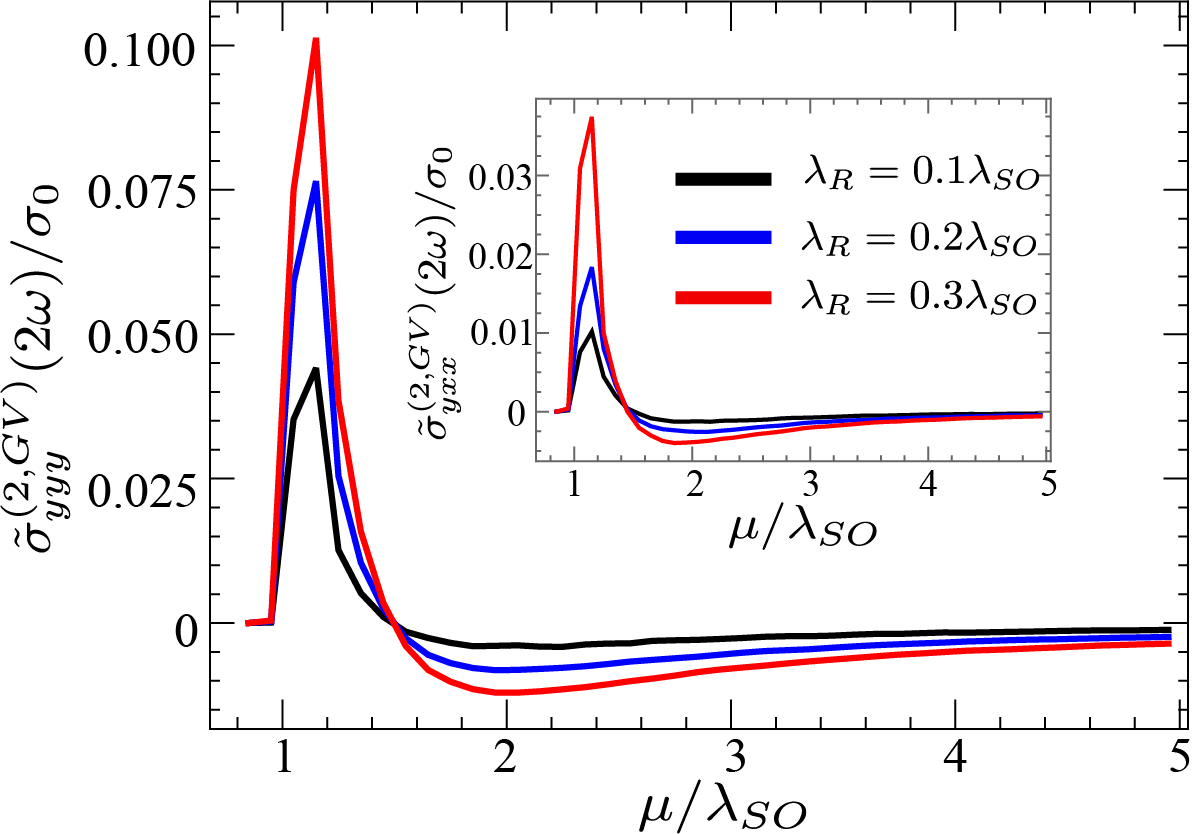}
\caption{Group velocity contribution to the second order intraband longitudinal $\tilde{\sigma}_{yyy}^{(2),GV}(2\omega)$ and (inset) transverse $\tilde{\sigma}_{yxx}^{(2),GV}(2\omega)$ electronic conductivities for $\Delta_{B}=0.5\lambda_{SO}$. The frequency dependence is captured in the normalization parameter ${\sigma_0}=e^3 v_F/\left[S\hbar(2\omega+i\Gamma)(\omega+i\Gamma)\right]$.}
\label{Fig2}
\end{figure}

Since the interplay between Rashba and Zeeman interactions does not affect the energy and spin distributions in the $x$ direction, all GV conductivity terms with an odd number of `$x$'s must vanish. For concreteness, in the following we will discuss the non-zero $\{\alpha \beta \gamma\}=\{yyy\}, \{yxx\}$ conductivity tensor components. The former (latter) describes the longitudinal (Hall) current generated in the system when the ac electromagnetic wave is applied orthogonal (parallel) to the static in-plane magnetic field. In both cases, the currents are perpendicular to the Zeeman field direction.
The conductivity tensors ${\tilde \sigma}_{yyy}^{(2,GV)}$ and ${\tilde \sigma}_{yxx}^{(2,GV)}$ are plotted as a function of chemical potential for a fixed frequency and $\Delta_B = 0.5\lambda_{SO}$, see Fig. \ref{Fig2}. Note that both conductivities show non-monotonic behavior with a maximum near $\mu_0\approx 1.11\lambda_{SO}$. The magnitude of this maximum increases for higher values of the Rashba coupling because this corresponds to larger asymmetries of the energy dispersion. The position of the maximum conductivity can be explained by closely looking at the energy dispersion of the conduction bands. When the chemical potential is below $\mu_0$, only the lowest conduction band contributes to the conductivity tensor, while above $\mu_0$, a second contribution appears from the highest conduction band. 
Each conduction band, however, gives contributions of opposite sign and the cumulative effect results in a decrease of the total conductivity for $\mu > \mu_0$.
For values of the chemical potential much larger than the spin-orbit coupling one notices that the conductivities approach zero. This is a consequence of the fact that more energetic charge carriers, which dominate the electronic transport in this regime, are not sensitive to the broken SIS and TRS that occur at lower energies near the bandgap. 

The physical origin of the GV contribution to a second order conductivity arises from the Rashba-Zeeman induced distortion of Fermi contours. Rashba coupling alone introduces spin-momentum locking in the system and the spins at opposite wave vectors are oriented in opposite directions (Fig. \ref{Fig1}c, second panel) and are equally populated.  This introduces a second order spin current in the system, {\it i.e.}, the spins oriented perpendicular to the electric field give a longitudinal spin current and spins oriented parallel (or anti-parallel) to the field generate a transverse spin current. The charge current, however, vanishes since the spins with opposite orientations move in opposite directions. Note that the spin current here refers to the spins with orientation along the plane of the sample. When a Zeeman field is introduced, in addition to the Rashba coupling,  it distorts the Fermi contours and creates population imbalance of opposite spins at opposite wave vectors (Fig. \ref{Fig1}c, last panel). This results in an asymmetry in the number of charges  moving in the opposite directions, and  leads to a second order nonzero charge current along both longitudinal and transverse directions. 

We refrain from showing the GV contribution to the spin and valley conductivities since those are zero for the case under consideration. Indeed, the $z$-polarized spin conductivity ${\tilde \sigma}_{s,\alpha\beta\gamma}^{(2,GV)}$ is nonzero only if the up-spin current and down-spin current are different. Because the Rashba-Zeeman interplay only creates asymmetry among the planar components of the spins, one can show that the second order current due to the up spins cancels the contribution from down spins, therefore resulting in ${\tilde \sigma}_{s,\alpha\beta\gamma}^{(2,GV)} = 0$. Regarding valley transport, although the asymmetric group velocity of the charge carriers individually leads to nonzero valley conductivities ${\tilde \sigma}_{\eta,\alpha\beta\gamma}^{(2,GV)}$ at $K$ and $K'$ Dirac points, their opposite value results in a perfect cancellation of the total valley conductivity. 

\subsection{Berry dipole contribution}
Next, we address the effect of Rashba-Zeeman interplay on charge, spin-polarized and valley-polarized Berry curvature profiles. The BD contribution  to the electronic, spin, and valley transport has an anomalous origin and may generate a second order transverse current when the incident ac field has a component perpendicular to the Zeeman magnetic field. For the Hamiltonian in consideration in Eq. \eqref{DiracHamiltonian}, we observe that only the second order spin-transport survives while the charge and valley transports vanish. Indeed, the Berry curvatures for up and down spins are the negative of each other and thus the total BD per Dirac cone vanishes, resulting in a zero second order conductivity. Similarly, the valley-polarized BD at $K$ and $K'$ are exactly zero, which prevents valley transport to take place. On the other hand, the spin-polarized Berry curvature takes the spin weight into account and the contributions from up and down spins add up. This results in a finite spin-polarized Berry dipole profile at a particular Dirac cone and a nonzero spin transport. 
\begin{figure}
\includegraphics[width=1.0\linewidth]{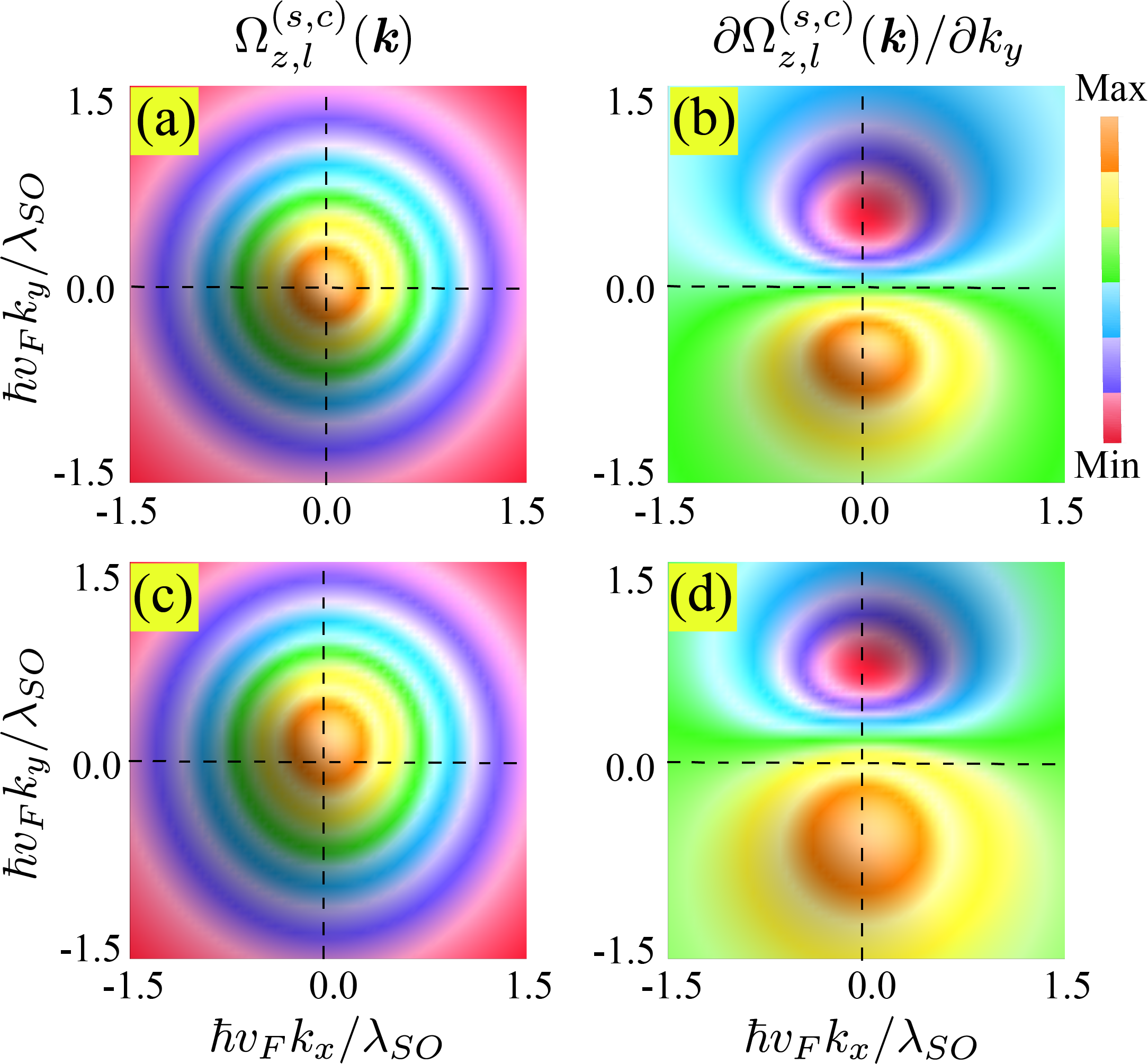}
\caption{(a) Momentum distribution of the spin-polarized Berry curvature $\Omega_{z,l}^{(s,c)}({\bm k})$ and (b) corresponding  derivative $\partial \Omega_{z,l}^{(s,c)}({\bm k})/\partial k_y$ for $\Delta_{B}=0$,  $\lambda_{R}=0$. Index `c' refers to contributions from the two conduction bands.
(c), (d) Same as (a) and (b) for  $\Delta_{B}=0.5\lambda_{SO}$, $\lambda_{R}=0.3\lambda_{SO}$. }
\label{Fig3}
\end{figure}

\begin{figure}
\includegraphics[width=1.0\linewidth]{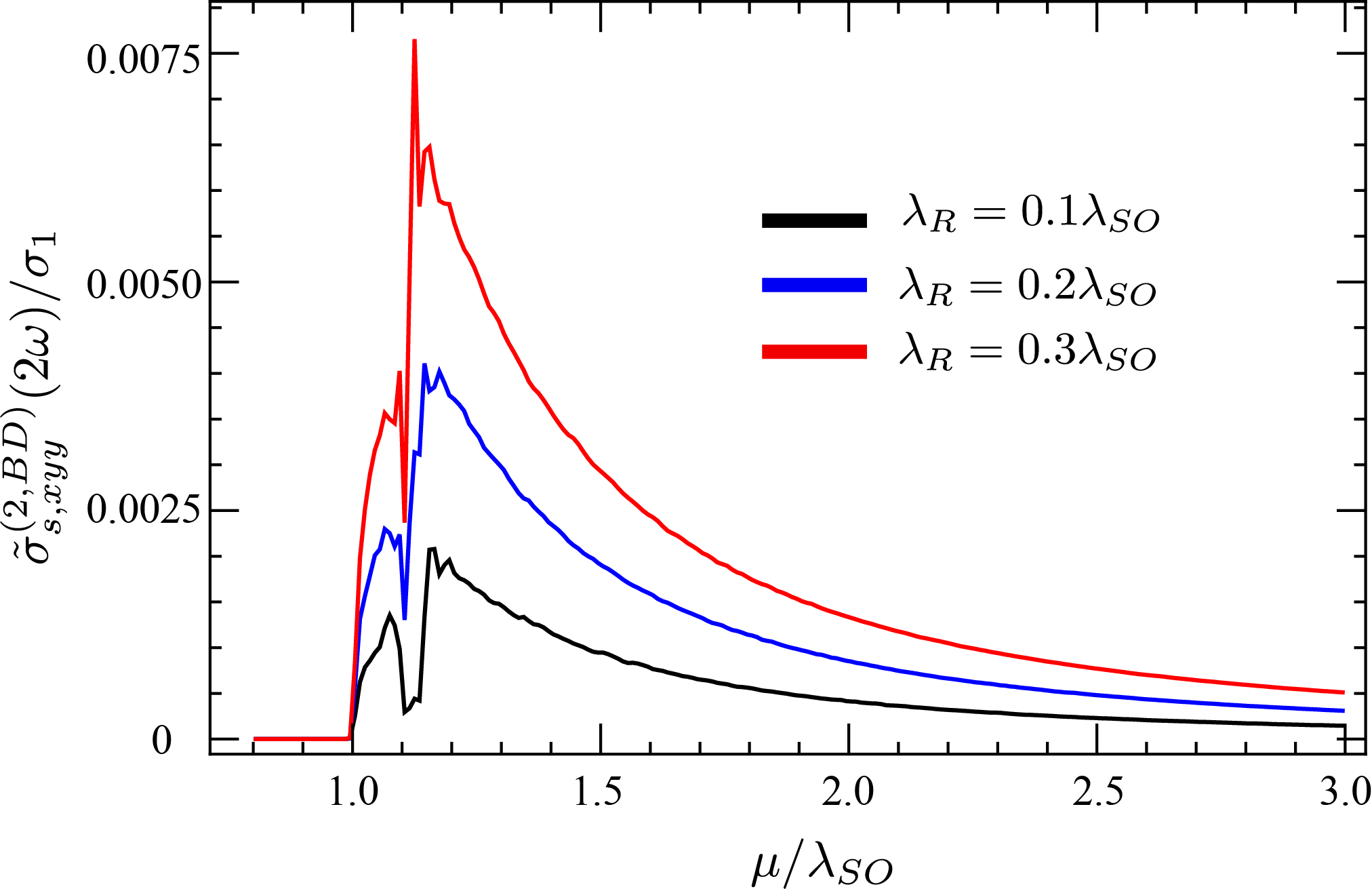}
\caption{Conductivities $\tilde{\sigma}_{s,xyy}^{(2),BD}(2\omega)$ for  $\Delta_{B}=0.5\lambda_{SO}$.
The frequency dependence is captured in the parameter ${\sigma_1}=e^3 \hbar v_F/\left[S(\omega+i\Gamma)\right]$. }
\label{Fig4}
\end{figure}

In Fig. \ref{Fig3} we plot the momentum dependence of spin-polarized Berry curvature profile $\Omega_{z,l}^{(s,c)}({\bm k})=\sum_{l=3,4}\Omega_{z,l}^{(s)}({\bm k})$ and its corresponding derivative  $\partial \Omega_{z,l}^{(s,c)}({\bm k})/\partial k_y$ for the conduction bands to demonstrate how they are affected by the Zeeman-Rashba coupling.
When both $\Delta_B$ and $\lambda_R$ are zero (panel \ref{Fig3}(a)) we notice that  spin-polarized Berry curvature is symmetric and its corresponding derivative is an odd function of $k_y$ (panel \ref{Fig3}(b)), therefore the Berry dipole moment vanishes. When the Rashba-Zeeman interplay is turned on, however,  it creates an asymmetry in the Berry curvature as shown in Fig. \ref{Fig3}c, where we observe that the center of the distribution is shifted along $k_y$-direction. As a result, the corresponding derivative in Fig. \ref{Fig3}d is no longer an odd function, and this guarantees that the integral in momentum space appearing in Eq. (\ref{berry-dipole}) is nonzero, and thus the BD moment is nonzero. As discussed before, the third contribution of (\ref{sigma2}), ${\tilde \sigma}_{s,xyy}^{(2,BDC)}$, follows the same principles as BD contribution and acts as a small correction to the BD contribution of the conductivity. In Fig. \ref{Fig4}, the contributions due to the BD to the spin-polarized conductivity $\tilde{\sigma}_{s,\alpha\beta\gamma}^{(2)}$ are shown for a fixed value of the Zeeman interaction ($\Delta_{B}=0.5\lambda_{SO}$) and for three values of Rashba coupling, as in Fig. \ref{Fig2}. The BD-enabled spin-conductivity  shows a non-monotonic behavior similar to that seen in the GV contribution term, and at large chemical potential the conductivity saturates near zero. We also observe a peculiar feature for chemical potential near $\mu_0$, which arises due to the numerical derivative taken near the crossing of two conduction bands.

\subsection{Interplay between Berry dipole moment  and tailored topological phase transitions}

\begin{figure}
\includegraphics[width=1.0\linewidth]{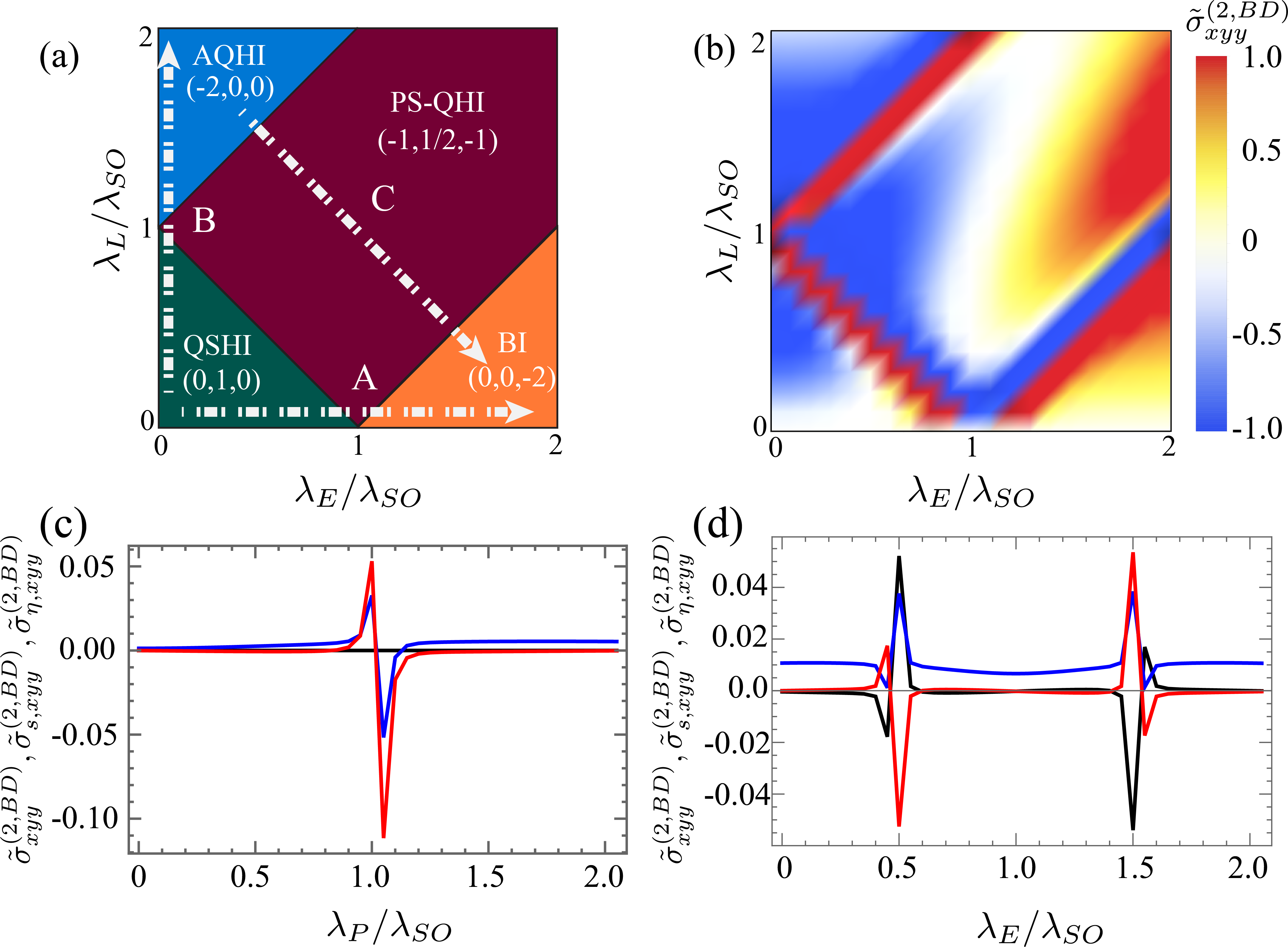}
\caption{(a) Topological phase diagram of the KM Hamiltonian without the Rashba and Zeeman terms. (b) The phase diagram of $\tilde\sigma_{xyy}^{(2,BD)}$ as a function of  $\lambda_{E}$ and $\lambda_{L}$ in the presence of Rashba-Zeeman couplings. The conducitvity is normalized with respect to the corresponding maximum value.
(c, d) Conductivities $\tilde\sigma_{xyy}^{(2,BD)}$ (black), $\tilde\sigma_{s,xyy}^{(2,BD)}$ (blue), and $\tilde\sigma_{\eta,xyy}^{(2,BD)}$ (red) (c) along paths A ($\lambda_P=\lambda_E$), B ($\lambda_P=\lambda_L$), and (d) C, as shown in panel (a).  Note that, for path B, the red color represents the charge conductivity and the black color represents the valley conductivity. The conductivities are evaluated for fixed values of $\Delta_B=0.5\lambda_{SO}$, $\lambda_{R}=0.3\lambda_{SO}$, and $\mu=5\lambda_{SO}$. The conductivities in (c, d) are normalized by $\sigma_1 S$.}
\label{Fig5}
\end{figure}

The Dirac half-gaps $\Delta_{\eta}^s$ in the Hamiltonian in Eq. (\ref{DiracHamiltonian}) can be externally tailored to host various topological phases. For example, for a 2D Kane-Mele topological insulator with two staggered sub-lattices a static electric field can be used to generate on-site potentials that will modify the gap as $\Delta_{\eta}^{s}=\eta s \lambda_{SO}-\lambda_{E}$, where $\lambda_E$ is the coupling with the field. Analogously, a high-frequency non-resonant circularly polarized laser can induce Floquet topological states in the monolayer, providing an additional knob for tailoring the energy  band structure via asymmetric valley coupling, {\it i.e.}, the Dirac gap change as $\Delta_{\eta}^{s}=\eta s \lambda_{SO}-\lambda_{E}-\eta \lambda_{L}$, where $\lambda_L$ represents the laser induced interaction. The topological phases accessible via these driving fields can be characterized by  topological invariants, namely the Chern ${\cal C}=\sum_{\eta,s} {\cal C}_s^{\eta}$, spin Chern  ${\cal C}_s=\sum_{\eta,s} s~ {\cal C}_s^{\eta}/2$, valley Chern  ${\cal C}_{\eta}=\sum_{\eta,s} \eta~{\cal C}_s^{\eta}$, and spin-valley Chern  $C_{s \eta}=\sum_{\eta,s} \eta s~{\cal C}_s^{\eta}/2$ numbers.  Here ${\cal C}_s^{\eta}=\eta~ \text{sign}[\Delta_s^{\eta}]/2$. By varying parameters $\lambda_E$ and $\lambda_L$, one can map a 2D topological phase diagram, which hosts electronic phases, including quantum spin Hall insulator (QSHI), band insulator (BI), anomalous quantum Hall insulator (AQHI), and polarized spin quantum Hall insulator (PS-QHI).  These phases are shown in Fig. \ref{Fig5}a and correspond to systems with all Dirac gaps open. Topological phase transitions take place at the borderlines between these phases (black solid curves) which mark the closing of at least one Dirac cone, while the two points $(\lambda_E/\lambda_{SO},\lambda_{L}/\lambda_{SO})=(1,0), (0,1)$ represent closing of two mass gaps. As we discuss next, the asymmetry introduced by modifying the Dirac gaps with external agents enables BD-induced charge, spin, and valley transport conductivities to be nonzero. 

For concreteness, here we will  focus on the conductivity tensor ${\tilde \sigma}_{xyy}^{(2,BD)}$, since it originates from the Berry curvature dipole moment and includes signatures of various topological properties of the system. In Fig. \ref{Fig5}b, we show the phase diagram of ${\tilde \sigma}_{xyy}^{(2,BD)}$ for fixed values of chemical potential $\mu=5\lambda_{SO}$, Zeeman interaction $\Delta_{B}=0.5\lambda_{SO}$, and Rashba coupling $\lambda_{R}=0.3\lambda_{SO}$. Away from the phase boundaries the conductivity shows nonlinear dependence on the magnitude of the Dirac gaps and is clearly seen in the PS-QHI phase, where the red and blue colors coexist, even though the sign of all the Dirac gaps remains the same. Near the topological phase transitions the conductivity is dominated by the smallest mass gap. The change of color near the phase boundaries reflects a change in Chern number, and a decrease in Chern number is associated with a color change from red to blue. This color change also marks a change in the Berry dipole direction. This behavior is consistent across all phase boundaries and can be better understood by comparing with the phase diagram in Fig. \ref{Fig5}a. Although the phase diagram of the conductivity demonstrates qualitative signatures of topological phase transitions and a decrease in Chern number, the magnitude of this change is difficult to estimate. 

A quantitative analysis of the second order transport properties across various phase transitions is shown in Fig. \ref{Fig5}c,d. There we plot the conductivities corresponding to charge-, spin-, and valley-currents along three paths in the phase diagram, namely: (A) QSHI$\rightarrow$AQHI  (B) QSHI$\rightarrow$BI (C) AQHI$\rightarrow$PS-QHI$\rightarrow$BI, as shown by marked arrows in Fig. \ref{Fig5}a. Along path (A), the Chern numbers  $\{ {\cal C}, {\cal C}_s, {\cal C}_{\eta}\}$ change as $\{0,1,0\}\rightarrow \{-2,0,0\}$, as we move from the QSHI to AQHI phase. Both charge and spin-polarized conductivities show similar behavior near the phase transition point $(\lambda_E/\lambda_{SO},\lambda_{L}/\lambda_{SO})=(0,1)$, with a resonance-antiresonance shaped behavior, indicating the change of Berry dipole direction across the phase transition.  The sign of Berry curvature near a particular Dirac cone is dependent on the sign of that Dirac gap. When one moves across a phase boundary the Dirac gap closes and reopens and this change in sign is reflected in the Berry curvature and an increase/decrease in the Chern number.  This sign change is subsequently transferred  to the Berry dipole. Furthermore, we observe that near the phase transitions the conductivities show strong enhancement in magnitude due to the fact that the inter-subband gap becomes small and  the Berry dipole increases as $1/|\Delta_{\eta}^{s}|^2$. Note that the valley conductivity vanishes along this path. The similarity between charge and spin conductivity is due to the fact that both charge and spin Chern numbers decrease in magnitude. Furthermore, we observe that the magnitude of the peaks across a phase transition for the charge conductivity is nearly two times that of the spin conductivity. This quantitative difference in the conductivities demonstrates that the charge- and spin-dipoles strongly respond to the topological phase transition and this response  
inherits the quantitative information about the change in Chern numbers. 

The fact that the conductivities and BD behavior inherit topology signatures near phase transitions can be seen across all the phase boundaries. For example, along path (B) $\lambda_{L}=0$ and the spin conductivity behavior remains the same as path (A), while the behavior of charge and valley conductivities  are switched. This is indeed in line with the fact that the change in charge Chern number across path (A) is equivalent to the change in valley Chern number across path (B). In order to further strengthen our argument, we also consider a more interesting path (C), where the system undergoes two phase transitions.  The charge Chern number increases across both boundaries, while the valley Chern number decreases across both boundaries. This is shown in Fig. \ref{Fig5}d, where the charge conductivity is the negative of the valley conductivity, and each conductivity shows the same behavior across both the boundaries. The spin Chern number, on the other hand, increases across the first boundary and then decreases across the second boundary. The corresponding conductivity behavior near the first boundary is opposite of the behavior at the second boundary. The above observations and arguments demonstrate clear signatures of topology in the conductivity near the phase transitions and could be extended to a three-dimensional topological phase diagram by adding another parameter to the mass term $\Delta_{\eta}^{s}$\cite{Malla2021}.

\begin{figure}
\includegraphics[width=1.0\linewidth]{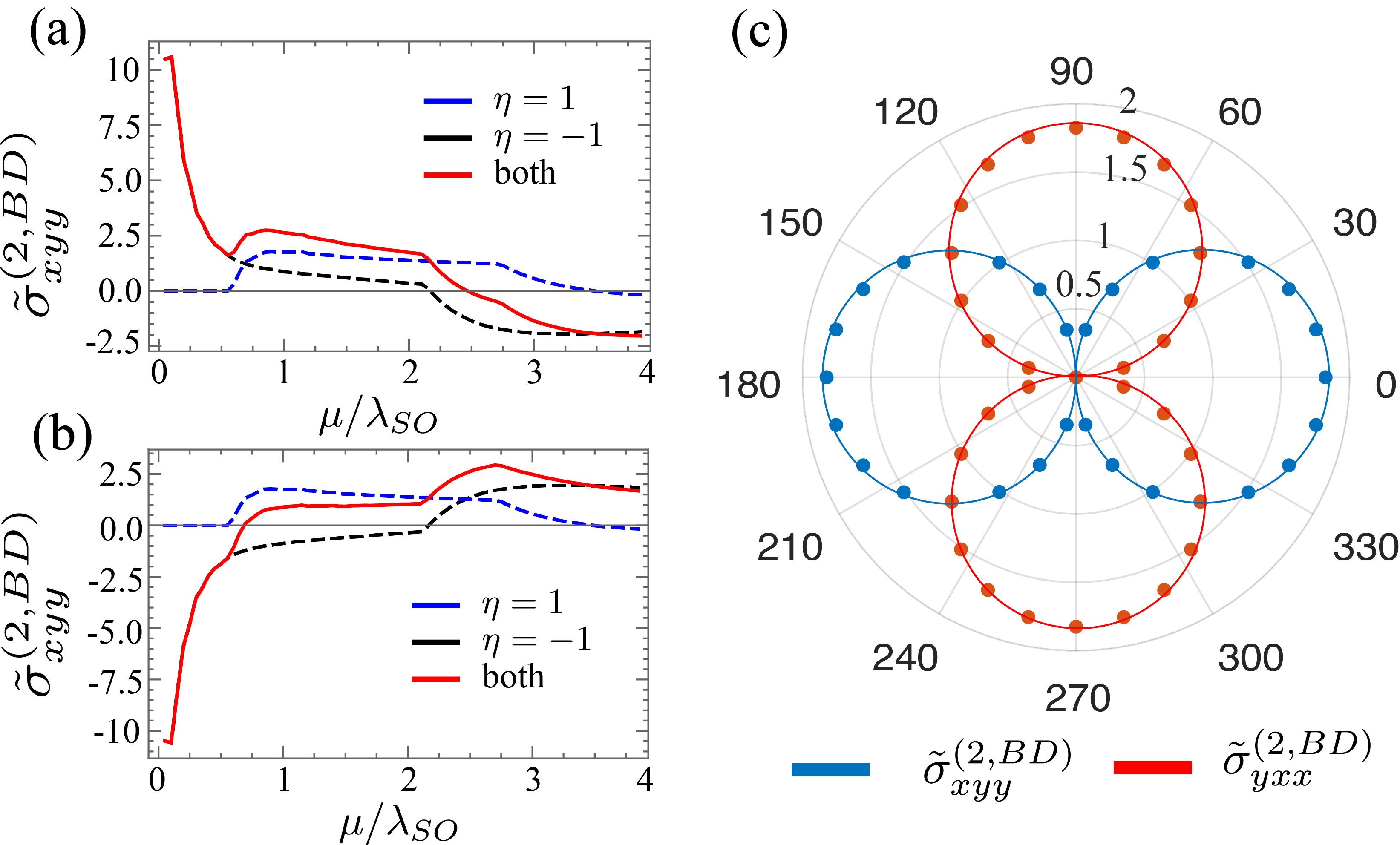}
\caption{(a, b) The $\mu$ dependence of the charge conductivities ${\tilde \sigma}_{xyy}^{(2,BD)}$ for $(\lambda_{E},\lambda_{L})/\lambda_{SO}=(0.3,1.5)$ (AQHI) and $(\lambda_{E},\lambda_{L})/\lambda_{SO}=(1.5,0.3)$ (BI), respectively. 
(c) The polar plot of $|{\tilde \sigma}_{xyy}^{(2,BD)}|$ (blue) and $|{\tilde \sigma}_{yxx}^{(2,BD)}|$ (red) are shown for the point in the AQHI phase as considered in (a).  The angle of the magnetic field is shown in degree. The conductivities are normalized by $\sigma_1 S $, $\mu=5\lambda_{SO}$.}
\label{Fig6}
\end{figure}

Next, we compare  ${\tilde \sigma}_{xyy}^{(2,BD)}$ evaluated at two different points in the phase diagram, $(\lambda_{E},\lambda_{L})/\lambda_{SO}=(0.3,1.5)$ (AQHI)
and $(\lambda_{E},\lambda_{L})/\lambda_{SO}=(1.5,0.3)$ (BI). These two points have the same band structure, however the signs of two of the mass gaps are different, and therefore they belong to different topological phases. To understand how the difference between the two phases affects the second order electronic conductivity of the system, we plot contributions from $K$ and $K'$ valleys separately (and the total conductivity) as a function of $\mu$ in Fig. \ref{Fig6}a (AQHI) and Fig. \ref{Fig6}b (BI). Note that, the contributions from $K$ valley are identical in both the phases, however that associated to the $K'$ valley for the AQHI phase is exactly the negative of the contribution in the BI phase. The sharp changes in the conductivity occur when $\mu$ coincides with one of the Dirac half-gaps $\Delta_{\eta}^{s}$. For small $\mu$, the conductivity only comes from the smallest gap, and the sign of that Dirac gap becomes equally important. Finally, in Fig. \ref{Fig6}c we plot the magnitudes of the conductivities $|{\tilde \sigma}_{xyy}^{(2,BD)}|$ and $|{\tilde \sigma}_{yxx}^{(2,BD)}|$ for the point considered in Fig \ref{Fig6}a as a function of the angle $\theta$ of the magnetic field as defined in Eq. \eqref{DiracHamiltonian}. The electric current is maximum when the electric field and the magnetic field are perpendicular to each other. The transverse nature of current suggests that the current flows parallel to the magnetic field (perpendicular to the electric field).  Note that, the magnitude of the conductivities corresponding to points  $(\lambda_{E},\lambda_{L})/\lambda_{SO}=(0.3,1.5)$ and $(\lambda_{E},\lambda_{L})/\lambda_{SO}=(1.5,0.3)$ are exactly equal, since they represent identical band structures. 

\section{Discussions and Conclusion}

In summary, we have showed that  massive Dirac fermions in centrosymmetric KM model can generate a second order transverse current when an out-of-plane symmetry broken Rashba coupling coexists with an in-plane magnetic field. We comment that the full KM tight-binding model also includes non-centrosymmetric part which in the low-energy regime  can be included via the trigonal warping term near the Dirac cones\cite{TRS8}.  A second order current due to this warping term has been demonstrated in various honeycomb systems\cite{warping1,warping2}. Comparing with our study, we note that the asymmetry in energy spectra created by Rashba Zeeman interplay is proportional to  $k_{\alpha}^{1/2}$, while the asymmetry created by the warping term is proportional to $k_{\alpha}^{3/2}$. So in a full KM model the Rashba-Zeeman induced second order term should dominate the optoelectronic response near the Dirac point, while the warping induced second order current dominates away from it. We also mention that a second kind of Rashba coupling ($\lambda_{R2}$) of intrinsic origin can appear in a typical KM material. Nevertheless, the strength of this coupling is proportional to the next-nearest-neighbor hopping parameter, therefore being often very small compared to $\lambda_{SO}$. Hence, we neglected it from our analysis.

In the quasi-static limit, we have demonstrated that the intraband current includes contributions from the asymmetry in the energy dispersion as well as from the asymmetry created in the Berry curvature. The GV contribution generates a second order current along longitudinal as well as the transverse direction, and remains normal to the Zeeman field. The BD contribution, however, only produces transverse current and is nonzero only when a component of the electric field is perpendicular to the Zeeman field. Since the transverse contributions from the GV and BD terms are perpendicular to each other, both contributions can be  experimentally probed separately. We briefly comment on the case of large frequency limit $\hbar\omega\sim \Delta_{\eta}^{s}$, where the interband conductivity ${\tilde \sigma}_{\alpha\beta\gamma}^{(1)}$ dominates. When $\mu$ lies in the middle of the band gap, interband transitions become the only contributing factor, but the second order nonlinear response in this case is zero in the absence of Rashba-Zeeman interplay \cite{Malla2021}.  When both Rashba and Zeeman couplings are present the asymmetry in the energy bands generates a resonance-induced second harmonic electric current perpendicular to the in-plane magnetic field. The material response then shows two spectral resonances corresponding to frequencies $\hbar\omega=2|\Delta_{\eta}^{s}|$ and $\hbar\omega=|\Delta_{\eta}^{s}|$. The resonance corresponding to $2\omega$ dominates since it resonates with a smaller gap. The Berry curvature contribution does not affect the interband current at the origin of the phase diagram, but becomes significant away from it.  

We also demonstrated that the topology of the low-energy Hamiltonian can be tailored via external parameters and the asymmetry between the Dirac gaps for different spins and valleys can play a significant role in the second order transverse current. We also show that the the Berry curvature induced dipole moment strength increases for Dirac cones with the smallest gap and the dipole changes sign across phase transitions. The recent synthesis of two-dimensional buckled members of graphene family, including silicene, germanene, stanene, and plumbene, as well as topologically non-trivial antiferromagnetic manganese chalcogenophosphates and perovskites provide a perfect platform to experimentally investigate second order Hall transport effects in Kane-Mele two-dimensional topological insulators. The ability to control each Dirac gap in these monolayers makes them a unique all-in-one material for  opto-electronics as well as spintronics based applications. 

\appendix

\section{Method}
We have followed the method prescribed in Ref.~\onlinecite{Mikhailov2016}, and included the electric field in the Hamiltonian as a scalar potential $\phi({\bm r},t)$. The equation of motion (\ref{EqMotRho}) is solved perturbatively in the energy eigen-basis $|\lambda\ket=|l{\bm k}\rangle$ and then the expression for the electric current is 
${\bm j}^{(n)}({\bm r}_0,t)=-e/2\sum_{ll'{\bm k}{\bm k}'}\bra l'{\bm k}'|\{\hat{{\bm v}} \delta({\bm r}_0-{\bm r})\}_{+}|l{\bm k}\ket \bra l{\bm k}|\hat{\rho}^{(n)}(t)|l'{\bm k}'\ket$ 
is used to evaluate the second order current. The first order solution leads to the Kubo formula for the linear conductivity. The second-order density matrix can be obtained from the first order density matrix and is expressed as
\begin{align}
&\bra\lambda|\hat{\rho}^{(2)}|\lambda'\ket_{t}=e^2 \int_{-\infty}^{\infty} d\omega_1 d\omega_2 e^{-i(\omega_1+\omega_2)t} \nonumber \\
&\hspace{20mm}\times \sum\limits_{\lambda''}\frac{ \bra\lambda|\phi_{\omega_2}({\bm r_2})|\lambda''\ket  \bra\lambda''|\phi_{\omega_1}({\bm r_1})|\lambda'\ket }{E_{\lambda'}-E_{\lambda}+\hbar(\omega_1+\omega_2+i\Gamma)} \nonumber \\
&\times 
\Bigg[ 
\frac{f_{\lambda'}-f_{\lambda''}}{E_{\lambda'}-E_{\lambda''}+\hbar(\omega_1+i\Gamma)}-\frac{f_{\lambda''}-f_{\lambda}}{E_{\lambda''}-E_{\lambda}+\hbar(\omega_2+i\Gamma)}
\Bigg],
\end{align}
and the corresponding expression for the second order current is given by 
\begin{multline}
{\bm j}^{(2)}({\bm r}_0, t)=-\frac{e^3}{2S} \int_{-\infty}^{\infty} d\omega_1 d\omega_2 \sum_{\tilde{q},{\bm q}_1,{\bm q}_2}
\phi_{{\bm q}_1\omega_1} \phi_{{\bm q}_2\omega_2}\\
\times
\sum_{\lambda,\lambda'}
\frac{\bra \lambda' | \{\hat{{\bm v}},e^{-i\tilde{{\bm q}}. {\bm r}_0}\}_+|\lambda\ket e^{i\tilde{{\bm q}}. {\bm r}_0-i (\omega_1+\omega_2)t}}{E_{\lambda'}-E_{\lambda}+\hbar(\omega_1+\omega_2+i\Gamma)}
\\
\times
\bra\lambda|e^{i{\bm q}_2. {\bm r}_2}|\lambda''\ket  \bra\lambda''|e^{i{\bm q}_1. {\bm r}_1}|\lambda'\ket \\
\times
\Bigg[ 
\frac{f_{\lambda'}-f_{\lambda''}}{E_{\lambda'}-E_{\lambda''}+\hbar(\omega_1+i\Gamma)}-\frac{f_{\lambda''}-f_{\lambda}}{E_{\lambda''}-E_{\lambda}+\hbar(\omega_2+i\Gamma)}
\Bigg]. 
\label{A2}
\end{multline}
Note that, the current does not change if we switch $(q_2,\omega_2,r_2)$ with $(q_1,\omega_1,r_1)$ in the second term of the last line in (\ref{A2}).

Now using the matrix element $\bra\lambda|e^{i{\bm q}_{1}. {\bm r}_1}|\lambda'\ket =\delta_{l,l'}-q_1 M_{l,l'}$, we can write the conductivity in a form which shows all types of virtual transitions involved in the second order process, and the expression reads
\begin{widetext}
\begin{multline}
{\bm j}^{(2)}({\bm r}_0, t)=-\frac{e^3}{2S} \int_{-\infty}^{\infty} d\omega_1 d\omega_2 \sum_{{\bm q}_1,{\bm q}_2}
\phi_{{\bm q}_1\omega_1} \phi_{{\bm q}_2\omega_2}e^{i({\bm q}_1+{\bm q}_2). {\bm r}_0-i (\omega_1+\omega_2)t}
\sum_{l,l',l'',{\bm k}}
\frac{\bra l',{\bm k} | \{\hat{{\bm v}},e^{-i({\bm q}_1+{\bm q}_2). {\bm r}_0}\}_+|l, \bm{k+q_1+q_2}\ket }{E_{\lambda'}-E_{\lambda}+\hbar(\omega_1+\omega_2+i\Gamma)}
\\
\times
\Bigg[ (\delta_{l,l''}-q_{2,\gamma}M_{l,l'',\bm{k+q_1}}) (\delta_{l'',l'}-q_{1,\beta}M_{l'',l',{\bm k}})
\frac{f_{l',{\bm k}}-f_{l'',\bm{k+q_1}}}{E_{l',{\bm k}}-E_{l'',\bm{k+q_1}}+\hbar(\omega_1+i\Gamma)}-
\\
(\delta_{l,l''}-q_{1,\beta}M_{l,l'',\bm{k+q_2}}) (\delta_{l'',l'}-q_{2,\gamma}M_{l'',l',{\bm k}})
\frac{f_{l'',\bm{k+q_2}}-f_{l,\bm{k+q_1+q_2}}}{E_{l'',\bm{k+q_2}}-E_{l,\bm{k+q_1+q_2}}+\hbar(\omega_1+i\Gamma)}
\Bigg].
\label{A3}
\end{multline}
\end{widetext}
The delta-functions in Eq. (\ref{A3}) represent virtual intraband transitions, while the terms with matrix elements $M_{l,l',{k_{\alpha}}}=2\hbar\langle v_{\alpha} \rangle_{l,l'}/(E_{l}-E_{l'})$ represent virtual interband transitions. We can separate terms depending on the type of virtual transitions, and we consider three cases, {\it (i)} $l=l'=l''$ (two intraband transitions) {\it (ii)} $l''=l\neq l'$ or $l''=l'\neq l$ (one intraband and one interband transitions) {\it (iii)} $l\neq l'\neq l''$ (two interband transitions). 

{\it Case i:} When virtual transitions occur within the same band only the delta-functions survive in Eq. (\ref{A3}), and we simplify the expression  inside the square bracket in (\ref{A3}) to

\begin{multline}
\frac{f_{l,{\bm k}}-f_{l,\bm{k+q_1}}}{E_{l,{\bm k}}-E_{l,\bm{k+q_1}}+\hbar(\omega_1+i\Gamma)}\\
-
\frac{f_{l,\bm{k+q_1}}-f_{l,\bm{k+q_1+q_2}}}{E_{l,\bm{k+q_1}}-E_{l,\bm{k+q_1+q_2}}+\hbar(\omega_1+i\Gamma)}, \nonumber
\end{multline}
using the identities $f_{l,k}-f_{l,k+q_1}\approx-q_{1\beta}\frac{\partial f_{lk}}{\partial k_{\beta}}$ and $\frac{\partial f_{l,k}}{\partial k_{\alpha}}-\frac{\partial f_{l,k+q_1}}{\partial k_{\alpha}}\approx-q_{1\beta}\frac{\partial^2 f_{lk}}{\partial k_{\alpha}\partial k_{\beta}}$ and arrive at a much simpler expression $\frac{q_{1\beta}q_{2\gamma}}{\hbar\omega_1+i\hbar\Gamma}\frac{\partial^2 f_{lk}}{\partial k_{\gamma}\partial k_{\beta}}$.  From here it is straightforward to find the GV contribution in Eq. (\ref{sigma2}).

{\it Case ii:} In the mixed case when the intraband transition and the interband transition coexist, the term in brackets in Eq. \eqref{A3}  for $l=l''$ becomes,
\begin{multline}
\frac{(-q_{1,\beta}M_{l,l',{\bm k}})(f_{l',{\bm k}}-f_{l,\bm{k+q_1}})}{E_{l',{\bm k}}-E_{l,\bm{k+q_1}}+\hbar(\omega_1+i\Gamma)}\\
-
\frac{(-q_{2,\gamma}M_{l,l',{\bm k}})(f_{l,\bm{k+q_1}}-f_{l,\bm{k+q_1+q_2}})}{\hbar(\omega_1+i\Gamma)}\nonumber
\end{multline}
and for $l'=l''$, 
\begin{multline}
(-q_{2,\Gamma}M_{l,l',\bm{k+q_1}})
\frac{f_{l',{\bm k}}-f_{l',\bm{k+q_1}}}{\hbar(\omega_1+i\Gamma)}\\
-
(-q_{1,\beta}M_{l,l',\bm{k+q_2}})
\frac{f_{l',\bm{k+q_2}}-f_{l,\bm{k+q_1+q_2}}}{E_{l',\bm{k+q_2}}-E_{l,\bm{k+q_1+q_2}}+\hbar(\omega_1+i\Gamma)}.\nonumber
\end{multline}
We can combine the two terms to arrive at the following expression
\begin{multline}
\frac{q_{1\beta}q_{2\Gamma}M_{l,l',k}}{\hbar\omega_1+i\hbar\Gamma} 
\Big[ 
\frac{\partial f_{l',k}-f_{l,k}}{\partial k_{\beta}}
\Big] \\
+q_{1,\beta}q_{2,\Gamma}\frac{\partial}{\partial k_{\Gamma}}\left[\frac{ M_{l,l',k}(f_{l',\bm{k}}-f_{l,\bm{k}})}{E_{l',\bm{k}}-E_{l,\bm{k}}+\hbar(\omega_1+i\Gamma)}\right].\nonumber
\end{multline}
The first term gives the BD contribution in Eq. (\ref{sigma2}). The second term can be expressed as a sum of two terms where one term contains the derivatives of the FD functions and the other term includes the remaining contributions including the FD functions. Note that, the term with the derivative of FD functions leads to the BDC term in (\ref{sigma2}), and the other term becomes the second part of Eq. (\ref{sigma1}).

{\it Case iii:} When both the virtual transitions are of interband type, the conductivity Eq. (\ref{A3}) needs no further simplification and we can simply put $q_1$ and $q_2$ to be zero and arrive at the first term of Eq. (\ref{sigma1}).

\section{Derivation of BD}
The expression for BD follows from the second term in  Eq. (\ref{sigma2})  and reads
\begin{multline}
{\tilde \sigma}_{\alpha\beta\gamma}^{(2,BD)}=
\frac{e^3\hbar}{S}\sum\limits_{l\neq l'{\bm k}}\frac{\bra \hat{v}_{\alpha}\ket_{l'l}}{\left[E_{l'l}+\hbar(\omega_1+\omega_2+i\Gamma)\right]}\\
\times\frac{\bra \hat{v}_{\gamma}\ket_{ll'}\frac{\partial f_{l'l}}{\partial k_{\beta}}}{E_{ll'}\hbar(\omega_1+i\Gamma)} .
\label{B1}
\end{multline}
In the semiclassical limit, $E_{l'l}\gg \hbar\omega_{1}, \hbar\omega_2$, we can neglect the frequencies in the denominator $E_{l'l}+\hbar(\omega_1+\omega_2+i\Gamma)$, and then Eq. (\ref{B1}) simplifies to
\begin{equation}
{\tilde \sigma}_{\alpha\beta\gamma}^{(2,BD)}=\frac{e^3\hbar}{S \hbar(\omega_1+i\Gamma)}\sum\limits_{l\neq l'{\bm k}}\frac{\bra \hat{v}_{\alpha}\ket_{l'l}\bra \hat{v}_{\gamma}\ket_{ll'}}{E_{l'l}^2}\frac{\partial f_{l'l}}{\partial k_{\beta}} .
\label{B2}
\end{equation} 
By using the fact that $\partial f_{l'l}/\partial k_{\beta}=-\partial f_{ll'}/\partial k_{\beta}$, and then separating the terms proportional to $\partial f_{l}/\partial k_{\beta}$ and 
 $\partial f_{l'}/\partial k_{\beta}$, we obtain
\begin{widetext}
\begin{equation}
{\tilde \sigma}_{\alpha\beta\gamma}^{(2,BD)}=\frac{e^3\hbar}{S \hbar(\omega_1+i\Gamma)}\sum\limits_{l'{\bm k}}\sum\limits_{l\neq l'}\frac{\left(\bra \hat{v}_{\alpha}\ket_{l'l}\bra \hat{v}_{\gamma}\ket_{ll'}\right)-\left(\bra \hat{v}_{\alpha}\ket_{l'l}\bra \hat{v}_{\gamma}\ket_{ll'}\right)^{*}}{E_{l'l}^2}\frac{\partial f_{l'}}{\partial k_{\beta}} .
\label{B3}
\end{equation} 
\end{widetext}
Equation (\ref{B3})  can be expressed as 
\begin{equation}
{\tilde \sigma}_{\alpha\beta\gamma}^{(2,BD)}=\frac{e^3\hbar}{S \hbar(\omega_1+i\Gamma)}\sum\limits_{l'{\bm k}}  \Omega_{\delta,l}({\bm k})\frac{\partial f_{l'}}{\partial k_{\beta}},
\label{B4}
\end{equation} 
where $\delta=\alpha \times \gamma$, is the direction of Berry curvature. For a 2D system, the Berry curvature vector always points in the $z$ direction and $\alpha$ and $\gamma$ must be linearly independent for a non-zero Berry curvature. Using integration by parts the summation in Eq. (\ref{B4}) can be separated into two parts, where the first part  is proportional to the FD functions at ${\bm k}\rightarrow \infty$ and the second part resembling Eq. (\ref{berry-dipole}). The FD function vanishes at  ${\bm k}\rightarrow \infty$ and Eq. (\ref{B4}) can be expressed as  
\be
D_{\alpha z}=\sum\limits_{l{\bm k}} \frac{\partial \Omega_{z,l}({\bm k})}{\partial k_{\alpha}} f_{lk}.
\label{B5}
\ee

\section*{Acknowledgements}
We acknowledge  the Laboratory Directed  Research  and  Development  program of  Los  Alamos  National  Laboratory  under  project  number 20190574ECR.  R.K.M. also thanks the Center for Nonlinear Studies at LANL for financial  support under project 20190495CR.

\bibliography{ref}

\end{document}